# Gradual Metaprogramming

Tianyu Chen
Indiana University
Bloomington, USA
chen512@iu.edu

Darshal Shetty
Indiana University
Bloomington, USA
dcshetty@iu.edu

Jeremy G. Siek
Indiana University
Bloomington, USA
jsiek@iu.edu

Chao-Hong Chen*
Meta
Menlo Park, USA
chaohong@meta.com

Weixi Ma*
Meta
Menlo Park, USA
mavc@meta.com

Arnaud Venet
Meta
Menlo Park, USA
ajv@meta.com

Rocky Liu
Meta
Menlo Park, USA
rockyliu4@meta.com

## Abstract

Data engineers increasingly use domain-specific languages (DSLs) to generate the code for data pipelines. Such DSLs are often embedded in Python. Unfortunately, there are challenges in debugging the generation of data pipelines: an error in a Python DSL script is often detected too late, after the execution of the script, and the source code location that triggers the error is hard to pinpoint.

In this paper, we focus on the scenario where a DSL embedded in Python (so it is dynamically-typed) generates data pipeline description code that is statically-typed. We propose gradual metaprogramming to (1) provide a migration path toward statically typed DSLs, (2) immediately provide earlier detection of code generation type errors, and (3) report the source code location responsible for the type error. Gradual metaprogramming accomplishes this by type checking code fragments and incrementally performing runtime checks as they are spliced together. We define MetaGTLC, a metaprogramming calculus in which a gradually-typed metalanguage manipulates a statically-typed object language, and give semantics to it by translation to the cast calculus MetaCC. We prove that successful metaevaluation always generates a well-typed object program and mechanize the proof in Agda.

***CCS Concepts:*** • **Software and its engineering → Semantics**; **Software development techniques**.

*Both authors contributed equally to this research.

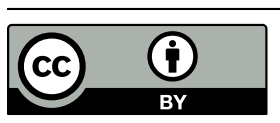

***Keywords:*** semantics, gradual typing, metaprogramming, data engineering, type-driven development, machine-checked proofs, Agda

## 1 Introduction

Data engineers use domain-specific languages (DSLs) to generate and manipulate the code for data pipelines. Examples include the Expression Language of Apache NiFi [Apache NiFi Team 2024], DAG files of Apache Airflow [Harenslak and De Ruiter 2021], the Jinja templating language used by Data Build Tool (dbt) [dbt Labs 2025], and the language of Ma et al. [2024]. Those DSLs are often embedded in Python. However, there are challenges in error detection: an error in a Python DSL script is often detected too late, after the execution of the script, and the source code location that triggers the error is hard to pinpoint.

We focus on the language of Ma et al. [2024], where a DSL embedded in Python (so it is dynamically-typed) generates data pipeline description code that is statically-typed. The goal of our research is to increase the data engineers' productivity in data pipeline construction and debugging. We propose *gradual metaprogramming*, which incrementally type checks code fragments as they are spliced together during metaevaluation and reports the source location of the problem when type checking fails. The goals are to enable finer-grained error detection and to improve the debugging of metaprograms that generate data pipelines. We define MetaGTLC, a metaprogramming calculus in which a gradually-typed metalanguage manipulates a statically-typed object language. The semantics of MetaGTLC is given by translation to its cast calculus MetaCC. We prove type safety for MetaGTLC, which means successful metaevaluation will always generate a well-typed object program. We mechanize MetaGTLC, MetaCC, and the type safety proof in Agda.

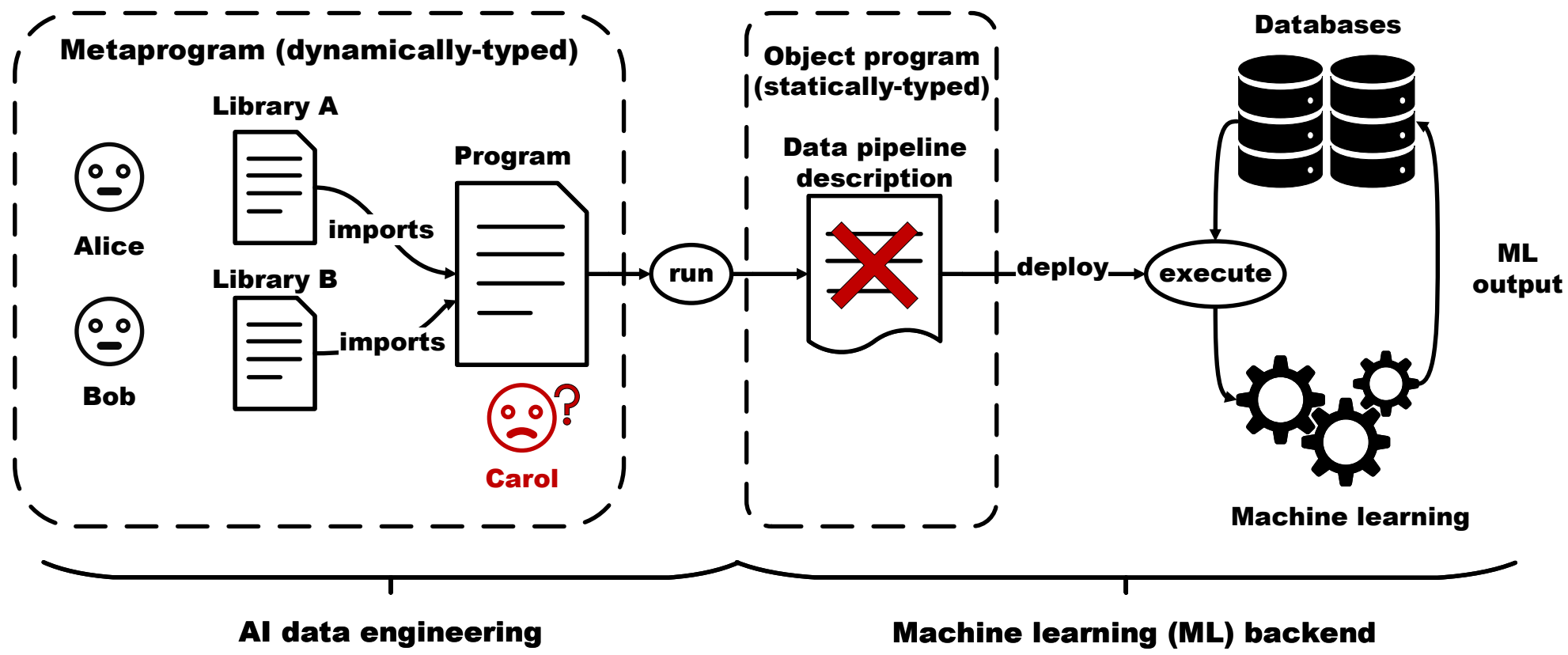

**Figure 1.** Difficulty in debugging dynamically-typed metaprograms that generate data pipelines

### 1.1 Generating Data Pipelines and Challenges in Error Detection

A data pipeline specifies the steps in which data is ingested, processed, and then stored [Crickard 2020; Harenslak and De Ruiter 2021; IBM 2024]. Data engineers generate and manipulate descriptions of data pipelines using DSLs, which are often embedded in Python. We begin by reviewing some metaprogramming systems that generate or manipulate data pipeline descriptions. Apache NiFi (https://nifi.apache.org/) is a generic framework of modeling the flow of data between systems. NiFi uses a metalanguage called the Expression Language to generate and manipulate the attributes and values of a FlowFile, which describes a piece of data that constitutes a data pipeline, such as a local file on the hard drive or a remote file on cloud storage. NiFi also supports manipulating data pipelines using Python scripts [Apache NiFi Team 2025]. After the data pipelines are generated, NiFi runs them on a Java virtual machine. Apache Airflow (https://airflow.apache.org/) is a tool that orchestrates different components for data processing in a data pipeline. The description for a data pipeline in Airflow is a directed acyclic graph (DAG). Airflow supports using "DAG files," a metalanguage that is Python augmented with additional metadata (e.g., when the pipeline should be executed) to dynamically generate and manipulate those DAGs. The Data Build Tool (https://www.getdbt.com/) uses Jinja, a templating language that translates to Python, to generate SQL database queries that construct data pipelines. Ma et al. [2024] design a language for AI data engineering. Their calculus is modeled on a real world DSL for feature engineering, a pre-processing step of machine learning that transforms raw data into a set of measurable properties. A surface language embedded in Python manipulates code in a statically-typed core calculus for data pipeline descriptions. When such a data pipeline description is deployed, it pulls data from databases, preprocesses it, and provides the pre-processed data as input to machine learning algorithms.

Data engineers face increasing challenges in debugging and pinpointing errors as metaprograms grow in size and complexity. Figure 1 demonstrates the difficulty in debugging a metaprogram written in a dynamically-typed DSL that generates statically-typed data pipeline descriptions (the scenario of Ma et al. [2024]). A team of three programmers, Alice, Bob, and Carol, are collaboratively constructing and debugging the same data pipeline. Carol constructs the metaprogram by invoking the code in Library A by Alice and Library B by Bob and adding code of her own. Carol then runs the metaprogram and generates a data pipeline description. Unfortunately, when Carol is ready to deploy the pipeline to the machine learning backend, she discovers (with confusion and disappointment) that the pipeline description is rejected by the typechecker and fails to compile. To make things worse, Carol has no clue about the source of the error. The mistake can be either in Alice's or Bob's library code, or in Carol's own code of the metaprogram.

### 1.2 Gradual Metaprogramming Helps Debugging Data Pipeline Generation

The main technical goal of our research is to speed up and streamline data engineers' debugging workflow of the Python metaprograms that generate data pipeline descriptions. We model data pipeline construction in a calculus, MetaGTLC, that combines two language features: (1) metaprogramming through quote and splice (2) gradual typing. The choice of performing metaprogramming through quote and splice is inspired by the rich literature on static type systems for metaprogramming, going back to the work on MetaML by Sheard and Taha [Sheard 1998; Taha and Sheard 1997, 2000] and Moggi [Moggi et al. 1999]. MetaML shows that it is possible to obtain a strong type safety guarantee: if the metaprogram successful type checks, then the code generated during metaevaluation will also type check. MetaML obtains this guarantee by tracking the type of a code fragment and by manipulates code fragments in a purely functional style using quote and splice.

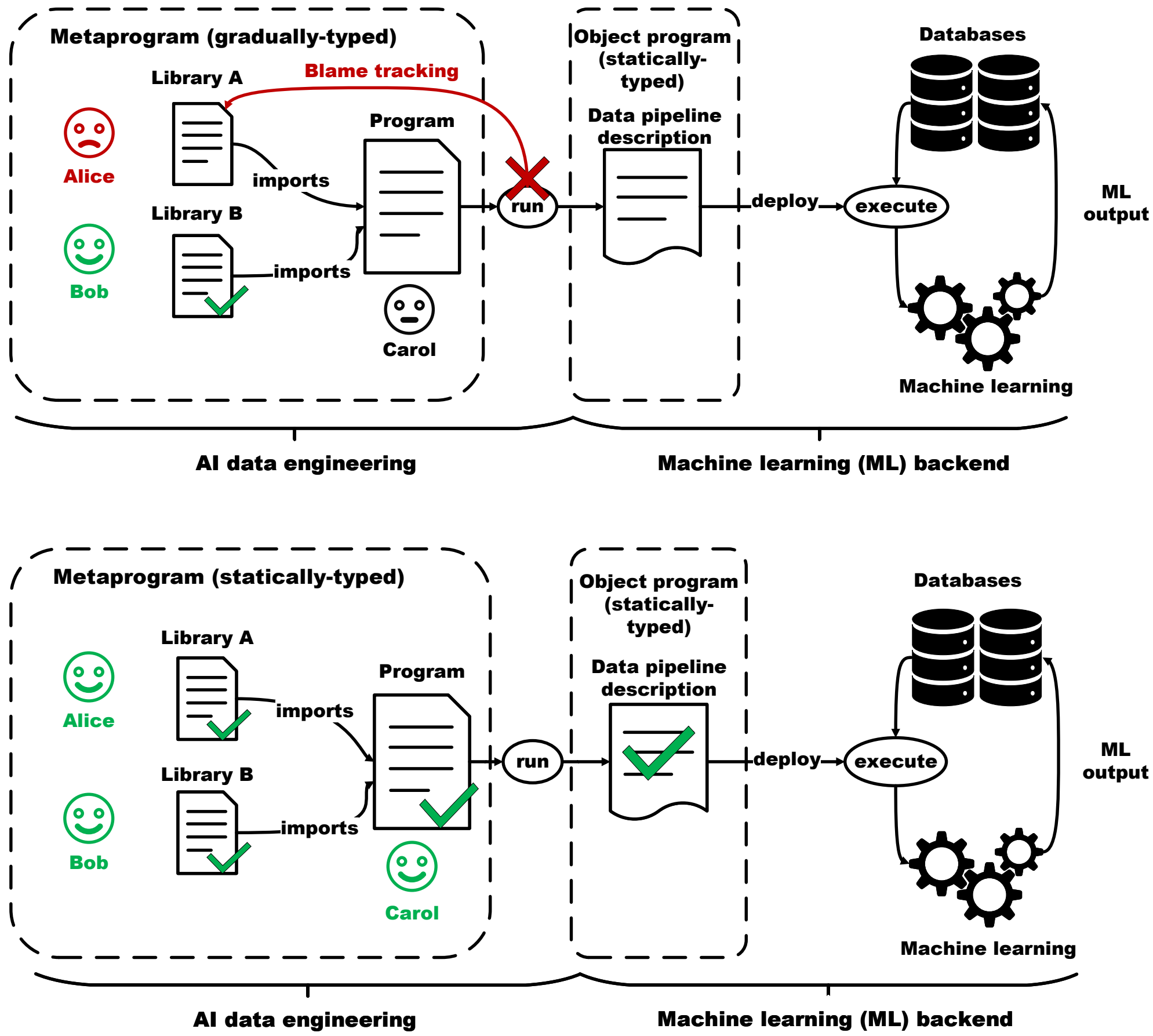

**Figure 2.** Our proposed workflow of generating data pipelines using gradual metaprogramming

Like MetaML, MetaGTLC also guarantees that generated data pipelines are well-typed. However, different from MetaML, whose metalanguage (i.e., ML) is statically typed, MetaGTLC employs a *gradually-typed* metalanguage. This design choice is based on the real-world needs of data engineers, who currently use Python for fast data pipeline development. As the metaprograms for data pipeline construction get increasing complicated and consist of more and more components, the data engineers will need to add type annotations by using a gradual typechecker for Python such as MyPy [Lehtosalo and mypy contributors 2025] so that the mistakes in the metaprograms are reported in earlier stages of the pipeline generation. Incrementally adding type information, the data engineers will eventually migrate to fully static metalanguage code that generates data pipelines. If that is the case, MetaGTLC guarantees the well-typedness of generated code fully statically just like MetaML. The calculus we study in this paper is agnostic to the details of the object language, just that it should be statically typed to model the statically-typed nature of the data pipeline description language of Ma et al. [2024]. As a result, we choose simply-typed lambda calculus (STLC) as our object language as a simple representative of statically-typed languages.

MetaGTLC speeds up debugging data pipeline construction because it reports bugs earlier and more precisely. In terms of time, MetaGTLC detects type errors either statically or during metaevaluation before the pipeline is even generated. In terms of code locations, through blame tracking, MetaGTLC pinpoints the metaprogram source code location responsible for the type error in its error message.

***Narrowing down the problem.*** MetaGTLC narrows the search for bugs in the metaprograms that generate data pipelines. The main problem of the error reporting of the language of Ma et al. [2024] is that a run of the metaprogram may generate an ill-typed pipeline description, and type errors are only detected when the pipeline description is type checked right before its deployment. To make matters worse, the generated data pipeline code could be in SQL or Apache Spark, which are completely different languages than the language (Python) that the programmers wrote the metaprogram in. In MetaGTLC, type errors are detected either statically or during metaevaluation, preventing ill-typed object programs from being generated, as is illustrated in Figure 2 (up). In practice, this approach provides two main benefits: first, it narrows the problem down to the generating program (in Python) and does not require the programmers to look at the generated object code (in SQL or Spark) at all. Second, it restricts the search for bugs to only part of the metaevaluation execution trace. The programmers of the AI data engineering team can then invest time and effort in adding type annotations to the code, making errors appear even earlier during metaevaluation. For example, Alice might discover a type error during compilation if she adds sufficient type annotations to Library A, which is great because the error is detected before metaevaluation even happens just like MetaML. Provided that Alice has fixed the bug in Library A, and the entire metaprogram becomes well-typed and runs correctly, the metaprogram is guaranteed to generate a well-typed data pipeline description. As is illustrated in Figure 2 (down), the pipeline description now correctly compiles and is ready for deployment, so everyone is smiling.

***Pinpointing the buggy code.*** MetaGTLC supports *blame tracking* [Findler and Felleisen 2002], thereby enabling modular runtime error messages. Consider the situation where Carol runs the metaprogram written in MetaGTLC and gets an error during metaevaluation. The *blame label* in the error message references the code fragment that is responsible for the error. In Figure 2 (up), the blame label points to a source code location inside Library A, so Carol is confident that neither her code nor Bob’s Library B is responsibility for the error. Carol could kindly ask Alice to fix Library A. Imagine another situation where Carol calls a function in Alice’s library with the wrong type. In that case, the blame goes to Carol, so Carol will have to fix her own code.

### 1.3 Gradual Metaprogramming to Improve Data Pipeline Generation

Gradual metaprogramming supports a seamless transition from dynamic to static, so it is backward compatible with existing code of a dynamically-typed DSL that generates data pipelines, while facilitating the migration towards static typing by enabling the incremental addition of type annotations. Furthermore, gradual metaprogramming is a first step towards accommodating richer typing disciplines in the future, such as information-flow control (IFC) for privacy-preserving machine learning.

***Migration and backward compatibility.*** MetaGTLC enables a seamless migration from dynamic metaprograms to static metaprograms through gradual typing. Gradual typing embeds dynamic typing, so existing dynamically-typed metaprograms and libraries will continue to run on MetaGTLC. The programmers can then incrementally add type annotations. Ideally, all libraries and metaprograms will eventually become statically typed (Figure 2, down).

**Table 1.** Typing paradigms of metaprogramming languages

| **Object language** | **Metalanguage** | | |
|---|---|---|---|
| | Static | Dynamic | Gradual |
| Static | MetaML/MetaOCaml<br>Template Haskell<br>Squid, T-LINQ | C++ Templates<br>C Preprocessor<br>Ma et al. | Miao & Siek<br>**MetaGTLC** |
| Dynamic | - | Lisp/Racket | - |
| Gradual | - | - | $\lambda^{G\circ}$ |

***Development productivity.*** MetaGTLC increases the development productivity of data engineers during both the initial prototyping phase and the follow-on debugging phase. Just like a dynamically-typed DSL (such as the language of Ma et al. [2024]), the programmers can leave out type annotations and make the metaprogram fully dynamically typed when they prototype the construction of a new data pipeline. However, unlike the language of Ma et al. [2024], MetaGTLC performs checking during metaevaluation, so ill-typed data pipelines are never generated. If the programmers encounter type errors during metaevaluation, which cause the pipeline generation to fail, they could add type annotations to the generating program gradually, thereby narrowing down the problem to only part of the metaevaluation execution.

***A Step Towards Privacy-Preserving AI..*** MetaGTLC is a first step towards combining metaprogramming with a gradual security type system for privacy-preserving machine learning. Recently, Chen and Siek [2024] design a gradual IFC calculus called $\lambda^{\star}_{\mathtt{IFC}}$, which satisfies both noninterference (the security guarantee) and the gradual guarantee. In $\lambda^{\star}_{\mathtt{IFC}}$, the programmer is free to choose when it is appropriate to increase the precision of type annotations and put in the effort to pass the static checks for higher performance, versus when it is appropriate to reduce the precision of type annotations, thereby deferring the enforcement to runtime for lower development cost. Similar to $\lambda^{\star}_{\mathtt{IFC}}$, MetaGTLC also uses coercions [Henglein 1994] to represent runtime checking. In future, we plan to extend MetaGTLC with IFC following the approach of $\lambda^{\star}_{\mathtt{IFC}}$.

### 1.4 Related Work

MetaGTLC is novel in the realm of metaprogramming languages because it fits in the less-explored design space of generating statically-typed object code using a gradually-typed metalanguage.

MetaML [Moggi et al. 1999; Sheard 1998; Taha and Sheard 2000]/MetaOCaml [Taha et al. 2004], Template Haskell [Sheard and Peyton Jones 2002], the Squid system of Scala [Parreaux et al. 2017], the T-LINQ language of constructing database queries [Cheney et al. 2013], C++ Templates [Abrahams and Gurtovoy 2004; Vandevoorde and Josuttis 2002; Veldhuizen 1995], Lisp [Steele 1990]/Racket [Flatt and PLT 2010], the language of Ma et al. [2024], and the calculus of Miao and Siek [2010] (which is adapted from the reflective metaprogramming calculus of Garcia and Lumsdaine [2009]) all provide multiple stages of computation, where the code of earlier stages can generate or manipulate code of later stages. We categorize these metaprogramming languages in Table 1 by the typing paradigms of their meta- and object languages. MetaML, MetaOCaml, and Template Haskell are statically-typed metalanguages that generate statically-typed code. Squid augments the quasiquotes of Scala with static type safety. The T-LINQ language models Microsoft's LINQ framework and statically guarantees the generation of well-formed SQL. Therefore, both Squid and T-LINQ also fall into the category of generating statically-typed object code from a statically-typed metalanguage. Lisp and Racket generate dynamically-typed object code from dynamically-typed metalanguages. The language of Ma et al. [2024], on the other hand, is similar to C++ Templates and the C preprocessor because it generates statically-typed code using a dynamically-typed metalanguage.

MetaGTLC is a gradually-typed metalanguage that generates statically-typed object language code. The only other language in this category is the calculus of Miao and Siek [2010]. Similar to the calculus of Miao and Siek [2010], MetaGTLC incrementally type checks the code fragments as they are spliced together during metaevaluation. Different from the calculus of Miao and Siek [2010], MetaGTLC takes a more standard approach to gradual metaprogramming. In particular, there is an explicit unknown type (here written $\star$) and runtime type checking happens at the boundaries of statically- and dynamically-typed code fragments during metaevaluation, analogous to the Gradually Typed Lambda Calculus (GTLC) of Siek and Taha [Siek and Taha 2006].

Parallel to the development of MetaGTLC, Yaguchi and Kameyama [2025] propose $\lambda^{G\circ}$, where a gradually-typed metalanguage generates gradually-typed object code. Gradual typing includes static typing, so they envision a gradually-typed metalanguage generating statically-typed code (the scenario of MetaGTLC) as one of the most promising applications of $\lambda^{G\circ}$. Compared to $\lambda^{G\circ}$, MetaGTLC supports the important feature of blame tracking, which aids the debugging of metalanguage code. Under the hood, MetaGTLC is based on the $\lambda$C calculus of [Siek et al. 2021] and uses coercions as the cast representation, while $\lambda^{G\circ}$ is based on the $\lambda^{G}$ calculus of Siek and Taha [2006] that uses type-based casts. We prefer coercions as the cast representation because coercions make it easier to support space efficiency and incorporate richer typing disciplines such as information-flow tracking for security in the future.

### 1.5 Technical Contributions

We have motivated how gradual metaprogramming could be useful for data pipeline generation by enabling earlier and more precise error detection while maintaining high development efficiency. The rest of the paper makes the following technical contributions:

- We design MetaGTLC, the first calculus for gradual metaprogramming using the standard approach to gradual typing.
- We define MetaCC, a cast calculus for gradual metaprogramming.
- We prove type safety for MetaGTLC, so a successful run of metaevaluation always generates well-typed object language code.
- We mechanize MetaGTLC, MetaCC, and the type safety proof in the Agda proof assistant.

Our Agda development is at the following link:

https://github.com/cty12/tyde2025-metagtlc-artifacts

## 2 Gradual Metaprogramming in Action

In this section, we demonstrate that gradual metaprogramming enhances the debugging experience for data pipeline construction because it (1) detects type error as metaevaluation happens (Section 2.1) and (2) enables pinpointing the source of the error through blame tracking (Section 2.2). In the examples, we generate statically typed database queries with SQL syntax using a gradually-typed metalanguage. Omitted type annotations on $\lambda$-abstractions, functions, and let-bindings of the metalanguage default to the statically-unknown type ($\star$).

### 2.1 Error Detection During Program Generation

MetaGTLC incrementally type-checks code fragments as they are spliced together. As a result, a programmer does not have to wait until data pipeline generation is completed for a type error to be detected. Instead, the error will be reported during the generation of the data pipeline. We shall discuss a simple example and contrast how errors would be caught by MetaGTLC versus in the language of Ma et al. [2024].

Consider the following program, which reads a number from a file and then uses that number to build a query that fetches records of people from the tperson table:

**Example 2.1** (Dynamically typed)**.** The type error is detected during metaevaluation in MetaGTLC.

```
1  let r = read_and_quote "i.txt" in
2  ≺SELECT * FROM tperson
3     WHERE age < (~r) + work≻
```

The metaprogram is dynamically typed, and the object code between the opening quote (≺) and closing quote (≻) is statically typed. The splice operator (∼) embeds metalanguage computation within object code. The `read_and_quote` function takes a filename and returns a quoted object language string. For example, if the file `i.txt` contains 42, the function call (`read_and_quote “i.txt”`) will return ≺ “42” ≻.

The programmer intends that the program should evaluate to a query that fetches people whose age is less than the sum of the number in the file and the number of years the person has worked at a job. However, there is a type mismatch. When run in the language of Ma et al. [2024], the error is not detected until after the metaprogram is finished, producing the following query:

```
SELECT * FROM tperson
WHERE age < "42" ✗ + work
```

This query does not typecheck because “42” is a string not an integer. The programmer discovers the type error only after the query is generated.

Gradual metaprogramming is able to detect the error earlier, during the generation of the query. The result of calling `read_and_quote` is bound to the variable $r$ without a type annotation, which means that $r$ is of the unknown type ($\star$). The unknown type is consistent with the type (`Code String`) returned by `read_and_quote`. The variable $r$ is then spliced into an object language database query. The spliced term has type `Int` because $x$ is an argument of integer addition. For the program to type check, $r$ is expected to be of type (`Code Int`), which is also consistent with its type $\star$. Even though the program type-checks because the typechecker permits values flowing from (`Code String`) to $\star$ and then to (`Code Int`), the program errors during metaevaluation. MetaGTLC adds casts between types that are consistent, and the casts get checked during metaevaluation. In the example, there is a cast from (`Code String`) to $\star$ when the return value of `read_c` is bound to $r$, which is followed by another cast from $\star$ to (`Code Int`) when $r$ is spliced into the quote. During metaevaluation, these two casts collide because `String` and `Int` are different types. In MetaGTLC, this type mismatch is detected during metaevaluation, thus preventing the ill-typed query from being generated.

We can detect the type mismatch even earlier by going fully static and annotating $r$ with (`Code Int`):

**Example 2.2** (Statically typed)**.** The type error is detected even earlier (when type-checking the metaprogram) if we add a type annotation (highlighted) on the let-binding.

```
1  let r: Code Int = read_and_quote "i.txt" in
2  ≺SELECT * FROM tperson
3     WHERE age < (∼r) + work≻
```

MetaGTLC detects the type mismatch when type-checking the metaprogram before metaevaluation starts. The typechecker rejects the program because `read_and_quote` returns (`Code String`) but the annotation expects (`Code Int`):

```
let r: Code Int = read_and_quote "i.txt" ✗ in
```

### 2.2 Blame Tracking for Pinpointing Errors

MetaGTLC improves the debugging efficiency of data pipeline construction because it is able to pinpoint the cause of a type error that happens during metaevaluation through blame tracking. Blame tracking is especially useful when the metaprogram consists of multiple modules written by different programmers, because the mistake may be in the source code of module A even though module B raises the error.

We consider a scenario where Carol is training a machine learning model that predicts whether a student would like to learn interactive theorem proving in Agda. She uses the number of lines of Haskell code and the number of lines of proofs that the student wrote in the past to compute the input to her model. Carol writes a metaprogram in MetaGTLC using the function `compose` from Alice's Library A and the function `sqr` from Bob's Library B. The metaprogram constructs database queries to the `tstudent` table. The `student` column of the table contains a string for the name of a student. The `haskell` and `proof` columns contain integers for the number of lines of Haskell code and proofs that the student wrote in the past, respectively.

The `compose` function from Alice's Library A accepts two functions, $f$ and $g$, as arguments and returns their composition in the object language:

```
1  /* Alice's Library A
2     compose has a static type signature */
3  compose : (Code Int -> Code Int)
4          -> (Code Int -> Code Int)
5          -> Code (Int -> Int)
6  compose f g = ≺ λx.∼(f (g ≺x≻))≻
```

The `sqr` function from Bob's Library B builds an expression in the object language that calculates the square of $x$:

```
1  /* Bob's Library B
2     sqr also has a static type signature */
3  sqr : Code Int -> Code Int
4  sqr x = ≺(∼x) * (∼x)≻
```

Both `compose` and `sqr` are annotated with fully static type signatures. The type signature of a function serves as the function's specification. If the library functions are invoked by a user with the wrong types, which violates the specification, the blame should go to the user instead of the library.

Carol's metaprogram constructs a database query that computes the input to the machine learning algorithm based on the number of lines of Haskell and proofs:

**Example 2.3** (Blame tracking)**.** MetaGTLC correctly assigns blame to Carol's code instead of Alice's or Bob's library.

```
1  /* Carol's program, which uses Library A, B
2     scale is dynamically typed */
3  scale x =
4    let r = int (read "i.txt") in
5    if r > 0 then ≺false≻
6             else ≺3 * (~x)≻
7
8  ≺SELECT
9    ((~(compose sqr scale)) (haskell + proof))
10   FROM tstudent≻
```

The program first defines a dynamically-typed `scale` function, which takes an argument $x$. The `scale` function branches on an integer parsed from the file `i.txt` and returns quoted terms of different types: if the integer is positive, `scale` returns quoted `false` (which is a bool); otherwise, it returns a quoted term of multiplying $x$ by 3 (which is an integer). Then, the program builds a database query that selects the value of applying the composition of `sqr` and `scale` to the sum of the number of lines of Haskell and proofs for every students in the table `tstudent`. We assume that the file `i.txt` contains 42.

In the language of Ma et al. [2024], metaevaluating Carol's program generates the following ill-typed query:

```
SELECT
((λx. false * false ✗) (haskell + proof))
FROM tstudent
```

The typechecker rejects the query because multiplication expects integers not `false`. On seeing the type error, Carol has little clue about the root cause of the problem: the mistake can be in either Carol's own code or code of the libraries. The error reporting of Ma et al. [2024] does not connect type errors in the generated query with source code locations in the metaprogram or the libraries that the metaprogram uses.

Through blame tracking, MetaGTLC is able to expose the root cause of the type error. The error message tells Carol that something is wrong with the function application *on line 9 of her program* (not Library A or Library B):

```
((~ (compose sqr scale) ✗) (haskell + proof))
```

The error message says that `compose` expects the return type of `scale` to be (`Code Int`), but `scale` returns a value of (`Code Bool`) during metaevaluation.

Under the hood, MetaGTLC automatically adds blame labels (source code location identifiers) to the runtime casts that it inserts during compilation. Later, when an error is raised, its blame label can point back to the corresponding source code location. In line 9 of Carol's code, `compose` expects the second argument to be of the function type (`Code Int` $\longrightarrow$ `Code Int`), but `scale` is of the unknown type ($\star$), so a cast in inserted at that location. At runtime this cast wraps `scale` in a proxy that checks whether the arguments and return values match the expected type. So when the function `compose` makes a function call to `scale` and `scale` returns ≺ `false` ≻, the proxy signals an error and blames the function application of `compose` on line 9 of Carol's code.

MetaGTLC enables modular error reasoning through blame tracking. A type signature serves as the specification, and a blame label identifies the code location to blame. In the example, even though the error is raised in the body of `compose`, a function in Library A by Alice, MetaGTLC tracks down the error and blames the correct location in Carol's code.

**Types and typing contexts**

| | | | |
|---|---|---|---|
| base types | $\iota$ | $\in$ | $\{\mathsf{Nat}, \mathsf{Int}, \mathsf{Bool}, \mathsf{Unit}\}$ |
| atomic types | $a$ | $\in$ | $\{\star, \iota\}$ |
| ground types | $G, H$ | $\in$ | $\{\iota, \star \longrightarrow \star, \mathsf{Code}\star\}$ |
| object language types | $S, T$ | $::=$ | $\iota \mid S \longrightarrow T$ |
| metalanguage types | $A, B$ | $::=$ | $\iota \mid A \longrightarrow B$ |
| | | $\mid$ | $\star \mid \mathsf{Code}\star \mid \mathsf{Code}\ T$ |
| typing contexts | $\Gamma$ | $::=$ | $\emptyset$ |
| | | $\mid$ | $\Gamma, x{:}\mathsf{otype}\ T$ |
| | | $\mid$ | $\Gamma, x{:}\mathsf{mtype}\ A$ |

$\boxed{A \sim B}$

$$\frac{}{\star \sim A} \qquad \frac{}{A \sim \star} \qquad \frac{}{\iota \sim \iota} \qquad \frac{A \sim A' \quad B \sim B'}{(A \longrightarrow B) \sim (A' \longrightarrow B')}$$

$$\frac{}{\mathsf{Code}\ T \sim \mathsf{Code}\ T} \qquad \frac{}{\mathsf{Code}\star \sim \mathsf{Code}\star}$$

$$\frac{}{\mathsf{Code}\star \sim \mathsf{Code}\ T} \qquad \frac{}{\mathsf{Code}\ T \sim \mathsf{Code}\star}$$

**Figure 3.** Types of MetaGTLC. Consistency between metalanguage types

# 3 Definition of MetaGTLC

In this section, we present the formal definition of MetaGTLC. We first define the types of MetaGTLC in Section 3.1. We then define the syntax and the type system of MetaGTLC in Section 3.2. Finally, we define the metaevaluation of MetaGTLC in Section 3.3.

## 3.1 Types of MetaGTLC

We define the types of MetaGTLC in Figure 3. The object language is statically typed, so a type of the object language ($T$) can be either a base type (`Nat`, `Int`, `Bool`, or `Unit`) or a function type. The metalanguage is gradually typed, so types of the metalanguage ($A$) include the statically-unknown type $\star$ (highlighted). The metalanguage also includes two types for quoted object code (highlighted): if the code is typed at T, the quoted code is of (`Code` $T$). If the quoted code goes through casts, its type may become `Code`$\star$, which means that the term represents some quoted code, but the type of the code is unknown. Ground types are types that are not $\star$ but can be cast into or out of $\star$, which include any base type ($\iota$), the simplest form of a function type that has unknown in its parameter and return type ($\star \to \star$), and the type for code of unknown type ($\texttt{Code}\star$). Atomic types only include base types and $\star$. (Atomic types are used when we define identity coercions in Section 4.1.) A typing context ($\Gamma$) is an association list that maps variables to their types. Each type is associated with a tag: if the type is of the object language, it has tag `otype`; otherwise if the type is of the metalanguage, it has tag `mtype`. The type consistency relation $A \sim B$ is used in the typing rules of MetaGTLC. Two types are consistent when they are equal except for the places where either type contains unknown type information.

### 3.2 Syntax and Type System of MetaGTLC

The syntax for MetaGTLC is defined in Figure 4. The object language is the simply-typed lambda calculus (STLC) with constants ($k$). The annotation term is to facilitate bidirectional type checking [Dunfield and Krishnaswami 2021] (we discuss the need for bidirectional typing later in this section). The code language is the object language (STLC) extended with *splice* (highlighted), which escapes to the computation of the metalanguage. The metalanguage is the gradually-typed lambda calculus (GTLC) extended with *quote* (highlighted), which produces a piece of code in the object language. If the type annotation on a metalanguage $\lambda$-abstraction is omitted, it defaults to $\star$. The term for splice in the object language and the term of function application in the metalanguage both carry *blame labels* ($\ell$). Those terms incur casts that may trigger cast errors during metaevaluation. When a cast error (blame) is reported, its blame label references the source code location responsible for that error.

The type system of MetaGTLC is also defined in Figure 4. The typing context $\Gamma$ maps a variable to its type, and we distinguish a variable in the metalanguage from one in the object language by looking at the type's tag (`mtype` or `otype`). The typing of the metalanguage takes the form $\Gamma \vdash^m M^m : A$. The typing rules extend the type system of GTLC with a rule for quote: rule $\vdash^m$-*quote* (highlighted) lifts the type of the quoted object code ($T$) to $\texttt{Code}\ T$.

The typing of the object language is bidirectional, taking the forms $\Gamma \vdash^o M^o \Rightarrow T$ (synthesis mode) and $\Gamma \vdash^o M^o \Leftarrow T$ (checking mode). The bidirectional typing rules extend those for STLC with one rule for splice (highlighted). We use bidirectional typing because it is a mechanism for inferring types that are needed during cast insertion. Consider the example:

**Example 3.1** (Splicing code of unknown type)**.**

```
λx . ≺1  +  ~x≻
```

The variable $x$ has type $\star$ but we can only splice fully-typed code into the object language (STLC). So MetaGTLC needs to insert a cast $c$:

Syntax of MetaGTLC

$$
\begin{array}{rlcl}
\text{object terms} & M^s & ::= & x \mid k \mid \lambda x.\, M^s \mid M^s\ M^s \mid M^s : T \\
\text{code terms} & M^o & ::= & x \mid k \mid \lambda x.\, M^o \mid M^o\ M^o \mid M^o : T \\
 & & \mid & \sim^{\ell}\ M^m \\
\text{metaterms} & M^m & ::= & x \mid k \mid \lambda x{:}A.\, M^m \mid (M^m\ M^m)^{\ell} \\
 & & \mid & \prec M^o \succ
\end{array}
$$

$\boxed{\Gamma \vdash^o M^o \Rightarrow T}$

$$
\vdash^o\text{-}\mathit{const}\ \frac{k : \iota}{\Gamma \vdash^o k \Rightarrow \iota}
\qquad
\vdash^o\text{-}\mathit{var}\ \frac{\Gamma(x) = \texttt{otype}\ T}{\Gamma \vdash^o x \Rightarrow T}
$$

$$
\vdash^o\text{-}\mathit{app}\ \frac{\Gamma \vdash^o L^o \Rightarrow (S \to T) \qquad \Gamma \vdash^o M^o \Leftarrow S}{\Gamma \vdash^o L^o\ M^o \Rightarrow T}
$$

$$
\vdash^o\text{-}\mathit{ann}\ \frac{\Gamma \vdash^o M^o \Leftarrow T}{\Gamma \vdash^o M^o : T \Rightarrow T}
$$

$\boxed{\Gamma \vdash^o M^o \Leftarrow T}$

$$
\vdash^o\text{-}\mathit{lam}\ \frac{(\Gamma, x{:}\texttt{otype}\ S) \vdash^o M^o \Leftarrow T}{\Gamma \vdash^o \lambda x.\, M^o \Leftarrow (S \to T)}
$$

$$
\vdash^o\text{-}\mathit{check\text{-}infer}\ \frac{\Gamma \vdash^o M^o \Rightarrow T}{\Gamma \vdash^o M^o \Leftarrow T}
$$

$$
\vdash^o\text{-}\mathit{splice}\ \frac{\Gamma \vdash^m M^m : A \qquad A \sim \texttt{Code}\ T}{\Gamma \vdash^o \sim^{\ell}\ M^m \Leftarrow T}
$$

$\boxed{\Gamma \vdash^m M^m : A}$

$$
\vdash^m\text{-}\mathit{const}\ \frac{k : \iota}{\Gamma \vdash^m k : \iota}
\qquad
\vdash^m\text{-}\mathit{var}\ \frac{\Gamma(x) = \texttt{mtype}\ A}{\Gamma \vdash^m x : A}
$$

$$
\vdash^m\text{-}\mathit{lam}\ \frac{(\Gamma, x{:}\texttt{mtype}\ A) \vdash^m M^m : B}{\Gamma \vdash^m \lambda x{:}A.\, M^m : (A \to B)}
$$

$$
\vdash^m\text{-}\mathit{app}\ \frac{\Gamma \vdash^m L^m : (A_1 \to A_2) \qquad \Gamma \vdash^m M^m : B \qquad A_1 \sim B}{\Gamma \vdash^m (L^m\ M^m)^{\ell} : A_2}
$$

$$
\vdash^m\text{-}\mathit{app}\star\ \frac{\Gamma \vdash^m L^m : \star \qquad \Gamma \vdash^m M^m : A}{\Gamma \vdash^m (L^m\ M^m)^{\ell} : \star}
$$

$$
\vdash^m\text{-}\mathit{quote}\ \frac{\Gamma \vdash^o M^o \Rightarrow T}{\Gamma \vdash^m \prec M^o \succ : \texttt{Code}\ T}
$$

**Figure 4.** Syntax and typing of the gradual metaprogramming calculus MetaGTLC

```
λx . ≺1  +  ~(x⟨ c ⟩))≻
```

But what should be the target type of the cast? Bidirectional typing provide a mechanism for inferring that type from the code that surrounds the splice. In this case, the addition operator expects an argument of type `Int`, so the target type of $c$ should be (`Code Int`). In general, Bidirectional typing uses two modes: (1) the normal synthesis mode in which the type of a term is determined by inspecting the term itself, and (2) checking mode, in which the context of a term specifies an expected type. This example shows that the typing for splice must be in checking mode so that the target type of the cast is provided by the context. (We show full cast insertion rules in Section 5.)

### 3.3 Metaevaluation of MetaGTLC

The metaevaluation of MetaGTLC is defined by the total function *meta-eval*. Metaevaluation may either result in an STLC term or a blame, or diverge, or get stuck:

$$\text{results} \qquad r ::= M^s \mid \texttt{blame}\ \ell \mid \texttt{diverge} \mid \texttt{stuck}$$

The `stuck` case is for the proof of type safety: we prove that metaevaluation never gets stuck in Theorem 6.6.

The *meta-eval* function takes a well-typed metalanguage term ($M^m$). The metaevaluation function first compiles $M^m$ into the cast calculus (intermediate representation) term $M^c$. The function then reduces $M^c$ using the reduction of MetaCC and reports the result of reduction. We define the cast calculus MetaCC in the next section and the compilation from MetaGTLC to MetaCC in Section 5.

**Definition 3.2** (Metaevaluation of MetaGTLC)**.** Let $M^m$ be a well-typed MetaGTLC term and suppose $M^m$ is compiled into the MetaCC term $M^c$: $\emptyset \vdash^m M^m : \texttt{Code}\ T \rightsquigarrow M^c$. The metaevaluation of $M^m$ is defined as:

$$
\begin{array}{lr}
\textit{meta-eval}\ M^m \triangleq M^s & \text{if } M^c \longrightarrow^* \prec M^s \succ \\
\textit{meta-eval}\ M^m \triangleq \texttt{blame}\ \ell & \text{if } M^c \longrightarrow^* \texttt{blame}\ \ell \\
\textit{meta-eval}\ M^m \triangleq \texttt{diverge} & \text{if } \forall L.\ M^c \longrightarrow^* L \\
 & \text{and } L \longrightarrow N \text{ for some } N \\
\textit{meta-eval}\ M^m \triangleq \texttt{stuck} & \text{otherwise}
\end{array}
$$

## 4 Definition of the Cast Calculus MetaCC

In this section, we define a cast calculus for gradual metaprogramming called MetaCC, which extends the $\lambda$C calculus of Siek et al. [2021] with quote and splice. A cast calculus (CC) is an intermediate representation where all casts are made explicit. We represent casts in MetaCC by defining coercions in Section 4.1. We then present the syntax and the type system of MetaCC in Section 4.2. Finally, we define the small-step operational semantics for MetaCC in Section 4.3. In the next section, we define a type-preserving compilation from MetaGTLC to MetaCC, so that the semantics of MetaGTLC is given by MetaCC.

### 4.1 Cast Representation Using Coercions

We present the syntax and typing of coercions in Figure 5. Coercions are combinators that specify the conversion between two types: $\vdash c : A \Rightarrow B$. The syntax and typing of identity, injection, projection, function, and sequence coercions are standard: an identity coercion ($\texttt{id}\ a$) goes from an atomic type ($a$, which is base or $\star$) to the same atomic type. An injection ($G!$) converts from a ground type ($G$) to $\star$, and a projection ($G?^{\ell}$) goes in the other direction, from $\star$ to a ground type. Projections are responsible for blame, so a projection carries a blame label ($\ell$). A function coercion ($c \rightarrow d$) consists of two sub-coercions: $c$ is responsible for casting the parameter type, and $d$ casts the return type. A sequence coercion ($c; d$) connects the sub-coercion $c$, which casts from $A$ to $B$, with the sub-coercion $d$, which casts from $B$ to $C$, and forms a coercion from $A$ to $C$.

Syntax of coercions (cast representation)

$$
\begin{array}{llcl}
\text{coercions} & c & ::= & \texttt{id}\ a \mid G! \mid G?^{\ell} \mid c \rightarrow d \mid c;d \\
 & & \mid & \texttt{code-id}\star \mid \texttt{code-id}\ T \\
 & & \mid & \texttt{code!}\ T \mid \texttt{code?}^{\ell}\ T
\end{array}
$$

$\vdash c : A \Rightarrow B$

$$\vdash\textit{-id}\ \frac{}{\vdash \texttt{id}\ a : a \Rightarrow a}$$

$$\vdash\textit{-inj}\ \frac{}{\vdash G! : G \Rightarrow \star} \qquad \vdash\textit{-proj}\ \frac{}{\vdash G?^{\ell} : \star \Rightarrow G}$$

$$\vdash\textit{-fun}\ \frac{\vdash c : B \Rightarrow A \qquad \vdash d : A' \Rightarrow B'}{\vdash c \rightarrow d : (A \rightarrow A') \Rightarrow (B \rightarrow B')}$$

$$\vdash\textit{-seq}\ \frac{\vdash c : A \Rightarrow B \qquad \vdash d : B \Rightarrow C}{\vdash c;d : A \Rightarrow C}$$

$$\vdash\textit{-code-id}\star\ \frac{}{\vdash \texttt{code-id}\star : \texttt{Code}\star \Rightarrow \texttt{Code}\star}$$

$$\vdash\textit{-code-id}T\ \frac{}{\vdash \texttt{code-id}\ T : \texttt{Code}\ T \Rightarrow \texttt{Code}\ T}$$

$$\vdash\textit{-code-inj}\ \frac{}{\vdash \texttt{code!}\ T : \texttt{Code}\ T \Rightarrow \texttt{Code}\star}$$

$$\vdash\textit{-code-proj}\ \frac{}{\vdash \texttt{code?}^{\ell}\ T : \texttt{Code}\star \Rightarrow \texttt{Code}\ T}$$

**Figure 5.** Syntax and typing of coercions

Compared with $\lambda$C, we introduce four new coercions between code types (highlighted in Figure 5). There are two identity coercions between code types: `code-id★` goes from `Code★` to `Code★`, and `code-id` $T$ goes from `Code` $T$ to `Code` $T$. A code injection coercion is similar to a regular injection, except that the former is between code types and goes from `Code` $T$ to `Code★`. A code projection coercion goes in the opposite direction, from `Code★` to `Code` $T$. Similar to a regular projection, a code projection is responsible for blame, so it also carries a blame label.

### 4.2 Syntax and Type System of MetaCC

The syntax and the type system of MetaCC are presented in Figure 6. All casts are made explicit in MetaCC. There is an explicit form for casts: $M^c\langle c \rangle$ (highlighted), where $c$ is the coercion to be applied to the MetaCC term $M^c$. If $M^c$ is typed at $A$ and the coercion $c$ goes from $A$ to $B$, then $M^c\langle c \rangle$ is typed at $B$ (rule $\vdash^c$-*cast*). In addition, we have a term `blame` $\ell$ (highlighted) for cast errors that may arise during metaevaluation. Each blame carries a blame label that identifies the source location of the error. A blame can have any type (rule $\vdash^c$-*blame*.)

Unlike MetaGTLC, terms for function application and splice in MetaCC no longer carry blame labels. During cast insertion, the blame labels go from the source code locations

Syntax of MetaCC

$$
\begin{array}{llcl}
\text{CC code terms} & M^{oc} & ::= & x \mid k \mid \lambda x.\, M^{oc} \mid M^{oc}\, M^{oc} \\
 & & \mid & M^{oc} : T \mid \sim M^{c} \\
\text{CC metaterms} & M^{c} & ::= & x \mid k \mid \lambda x{:}A.\, M^{c} \mid M^{c}\, M^{c} \\
 & & \mid & \prec M^{oc} \succ \\
 & & \mid & M^{c}\langle c \rangle \mid \texttt{blame}\ \ell
\end{array}
$$

$\Gamma \vdash^{oc} M^{oc} : T$

$$
\vdash^{oc}\textit{-const}\ \frac{k : \iota}{\Gamma \vdash^{oc} k : \iota} \qquad \vdash^{oc}\textit{-var}\ \frac{\Gamma(x) = \texttt{otype}\ T}{\Gamma \vdash^{oc} x : T}
$$

$$
\vdash^{oc}\textit{-lam}\ \frac{(\Gamma, x{:}\texttt{otype}\ S) \vdash^{oc} M^{oc} : T}{\Gamma \vdash^{oc} \lambda x.\, M^{oc} : (S \rightarrow T)}
$$

$$
\vdash^{oc}\textit{-app}\ \frac{\Gamma \vdash^{oc} L^{oc} : (S \rightarrow T) \qquad \Gamma \vdash^{oc} M^{oc} : S}{\Gamma \vdash^{oc} L^{oc}\, M^{oc} : T}
$$

$$
\vdash^{oc}\textit{-ann}\ \frac{\Gamma \vdash^{oc} M^{oc} : T}{\Gamma \vdash^{oc} (M^{oc} : T) : T}
$$

$$
\vdash^{oc}\textit{-splice}\ \frac{\Gamma \vdash^{c} M^{c} : \texttt{Code}\ T}{\Gamma \vdash^{oc} \sim M^{c} : T}
$$

$\Gamma \vdash^{c} M^{c} : A$

$$
\vdash^{c}\textit{-const}\ \frac{k : \iota}{\Gamma \vdash^{c} k : \iota} \qquad \vdash^{c}\textit{-var}\ \frac{\Gamma(x) = \texttt{mtype}\ A}{\Gamma \vdash^{c} x : A}
$$

$$
\vdash^{c}\textit{-lam}\ \frac{(\Gamma, x{:}\texttt{mtype}\ A) \vdash^{c} M^{c} : B}{\Gamma \vdash^{c} \lambda x{:}A.\, M^{c} : (A \rightarrow B)}
$$

$$
\vdash^{c}\textit{-app}\ \frac{\Gamma \vdash^{c} L^{c} : (A \rightarrow B) \qquad \Gamma \vdash^{c} M^{c} : A}{\Gamma \vdash^{c} L^{c}\, M^{c} : B}
$$

$$
\vdash^{c}\textit{-quote}\ \frac{\Gamma \vdash^{oc} M^{oc} : T}{\Gamma \vdash^{c} \prec M^{oc} \succ : \texttt{Code}\ T}
$$

$$
\vdash^{c}\textit{-cast}\ \frac{\Gamma \vdash^{c} M^{c} : A \qquad \vdash c : A \Rightarrow B}{\Gamma \vdash^{c} M^{c}\langle c \rangle : B}
$$

$$
\vdash^{c}\textit{-blame}\ \frac{}{\Gamma \vdash^{c} \texttt{blame}\ \ell : A}
$$

**Figure 6.** Syntax and typing of the cast calculus MetaCC

in MetaGTLC terms into coercions (specifically, projections and code projections) in MetaCC terms.

Compared with the bidirectional typing of the code terms in MetaGTLC, the typing for code terms is uni-directional in MetaCC. This is because all types are already recorded inside coercions in MetaCC. Also, type consistency in MetaGTLC turns into type equality in MetaCC, because all implicit type conversions are made explicit after cast insertion.

### 4.3 Operational Semantics of MetaCC

We define the small-step operational semantics for the cast calculus MetaCC in Figure 7. The semantics of MetaCC is based on that of $\lambda$C, but augmented with quote and splice for metaprogramming.

Similar to $\lambda$C, values in MetaCC include constants, $\lambda$-abstractions, and values wrapped with inert (value-forming) coercions. In addition, quoted STLC terms (that is, there is no splice inside) are also values. MetaCC is a call-by-value calculus, and the order of evaluation is from left to right.

The reduction of the cast calculus contains two relations: the metalanguage reduction takes the form $M^c \longrightarrow N^c$, where $M^c, N^c$ are metaterms of the cast calculus, and the code language reduction takes the form $M^{oc} \longrightarrow_o N^{oc}$, where $M^{oc}, N^{oc}$ are code terms of the cast calculus. When reducing a quoted code term, the metalanguage reduction resorts to the code language reduction (rule *$\xi$-quote*). On the other hand, when reducing a spliced term, the code language reduction turns to the metalanguage reduction (rule *$\xi$-splice*). A frame is a non-recursive evaluation context. Reduction under frames is grouped in the congruence ($\xi$) rules, where if the inner term takes a step, the term produced by plugging the inner term into a frame also takes a step.

Many metalanguage reduction rules are standard and mirror those in the $\lambda$C calculus. The $\beta$ rule is standard for a call-by-value calculus. The *fun-cast* rule distributes a function coercion onto its argument and return value. An identity coercion simply goes away (rule *id*). A sequence coercion is split into two consecutive coercions (rule *seq*). A pair of injection and projection goes away if the source of the injection is identical to the target of the projection (rule *proj*); otherwise, a cast error is signaled, blaming the projection (rule *proj-blame*).

MetaCC includes four rules that handle coercions between code types (highlighted). The rules for identity coercions between code types, *code-id*★ and *code-idT*, are analogous to rule *id*: the identity coercions on Code★ or Code $T$ simply go away. Rules *code-proj* and *code-proj-blame* are analogous to *proj* and *proj-blame* but for code types: if source and target types are the same, the pair of injection and projection goes away; otherwise, the code projection is blamed.

The *splice* rule follows the *splice* rule of the metaprogramming calculus of Garcia and Lumsdaine [2009]: the quoted object language term (which no longer contains splice) is spliced in.

There are three rules that propagate blames during metaevaluation: *$\xi$-blame*, *quote-splice-blame*, and *$\xi$-splice-blame* (highlighted). Rule *$\xi$-blame* propagates a blame through a frame of the metalanguage. Rules *quote-splice-blame* and *$\xi$-splice-blame* work in conjunction: metaevaluation inside a splice may error, so *$\xi$-splice-blame* lifts a spliced blame out of a code language frame; when the spliced blame reaches a quote, the pair of splice and quote goes away, and it reduces to just a blame by rule *quote-splice-blame*.

As usual, the multi-step reduction of MetaCC is defined as the reflexive transitive closure of single-step reduction:

$$
\frac{}{M^c \longrightarrow^* M^c} \qquad \frac{L^c \longrightarrow M^c \qquad M^c \longrightarrow^* N^c}{L^c \longrightarrow^* N^c}
$$

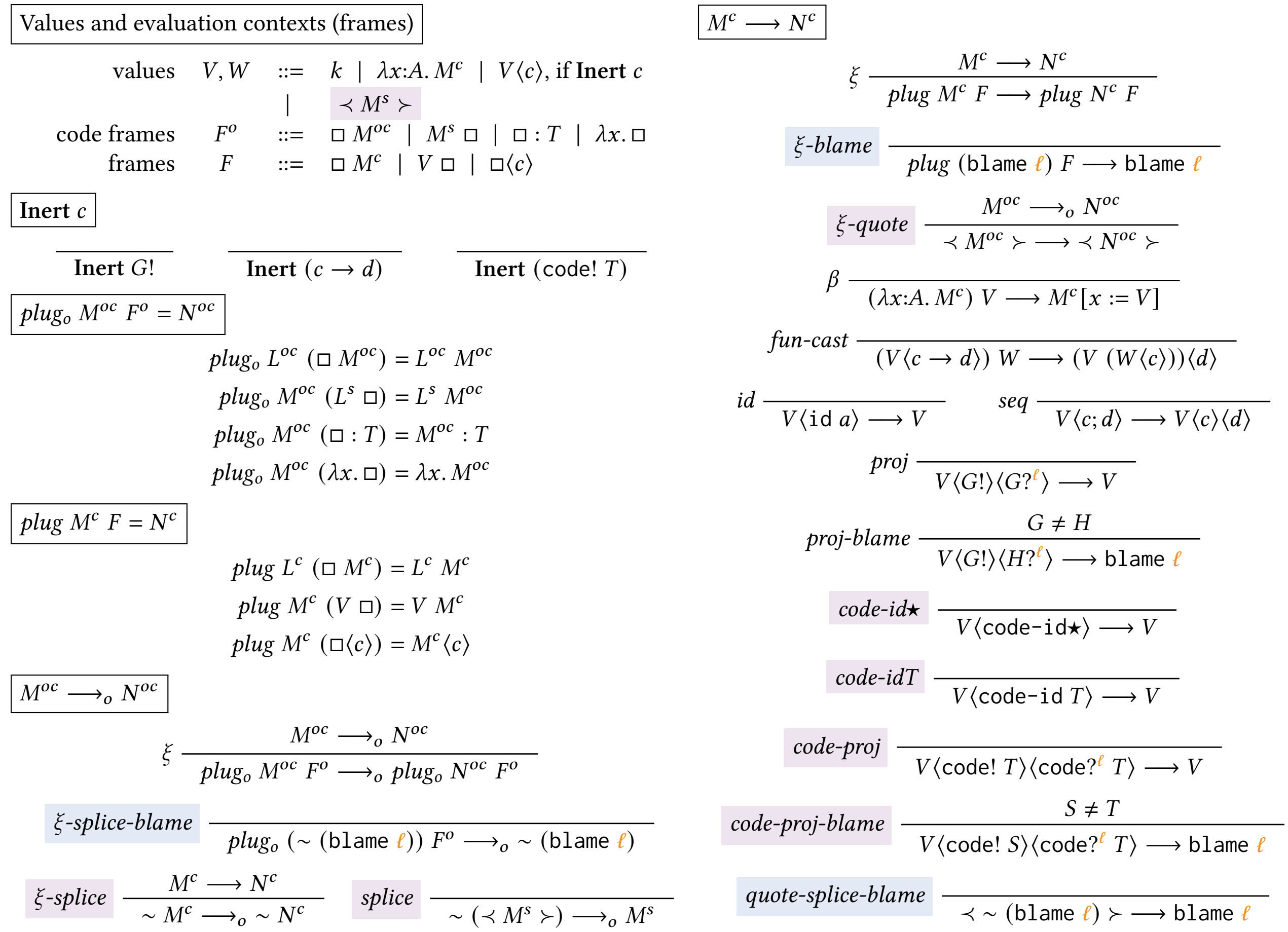

**Figure 7.** Small-step operational semantics of MetaCC

## 5 Compilation From MetaGTLC to MetaCC

In this section, we define the compilation from the gradual metaprogramming language MetaGTLC to its cast calculus MetaCC by inserting casts.

We first define the coerce function that takes two types that are consistent ($A \sim B$) as well as a blame label and generate a coercion that casts from $A$ to $B$ (Figure 8). A cast from $\star$ to a type that is not $\star$ (or the other way around) will always route through the corresponding ground type. We define a helper function *ground*, which takes a type $A$ that is not $\star$ and returns the corresponding ground type $G$ that is consistent with $A$.

The compilation from MetaGTLC to MetaCC is type-directed (Figure 9). Most rules are straightforward and recursively compile the sub-terms. We discuss the interesting rules for function application and splicing.

***Function application.*** If the function is typed at $(A_1 \to A_2)$, we insert one coercion on the argument that goes from that argument's type ($B$) to the argument of the function ($A_1$) after we recursively compile the function and its argument (rule $\rightsquigarrow^m$-*app*). Otherwise, if the type of the function is unknown ($\star$), in addition to the cast on the argument that goes from that argument's type to $\star$, we insert another cast from $\star$ to the ground function type on the function being applied (rule $\rightsquigarrow^m$-*app*$\star$). In both cases, the blame labels on the inserted casts come from the function application.

***Splicing.*** In rule $\rightsquigarrow^o$-*splice*, we first recursively compile the metalanguage subterm $M^m$. We need to insert a coercion from the type of $M^m$ (which is $A$) to Code $T$. The type $T$ comes from type checking the splice. Recall that in Section 3.2, we demonstrated using an example that bidirectional typing is required so that splice is always in checking mode to provide the target type of the inserted cast. The blame label of the inserted coercion comes from the syntax of splice. Finally, we splice the compiled sub-term with the inserted coercion.

## 6 Type Safety for MetaGTLC

In this section, we prove that successful metaevaluation in MetaGTLC always produces well-typed object code in STLC (Theorem 6.6). The proof depends on "progress" (Lemma 6.1) and "preservation" (Lemma 6.2) of the cast calculus MetaCC.

$$\boxed{\textit{ground}\ A = G}$$

$$\begin{aligned}
\textit{ground}\ \iota &= \iota \\
\textit{ground}\ (A \to B) &= \star \to \star \\
\textit{ground}\ (\texttt{Code}\ T) &= \texttt{Code}\star \\
\textit{ground}\ \texttt{Code}\star &= \texttt{Code}\star
\end{aligned}$$

$$\boxed{\textit{coerce}\ A\ B\ \ell = c}$$

$$\begin{aligned}
\textit{coerce}\ \iota\ \iota\ \ell &= \texttt{id}\ \iota \\
\textit{coerce}\ \star\ \star\ \ell &= \texttt{id}\ \star \\
\textit{coerce}\ \star\ G\ \ell &= G?^{\ell} \\
\textit{coerce}\ G\ \star\ \ell &= G! \\
\textit{coerce}\ \star\ A\ \ell &= (\textit{coerce}\ \star\ G\ \ell); (\textit{coerce}\ G\ A\ \ell) \\
&\quad \text{where } G = \textit{ground}\ A \\
\textit{coerce}\ A\ \star\ \ell &= (\textit{coerce}\ A\ G\ \ell); (\textit{coerce}\ G\ \star\ \ell) \\
&\quad \text{where } G = \textit{ground}\ A \\
\textit{coerce}\ (A \to B)\ (C \to D)\ \ell &= (\textit{coerce}\ C\ A\ \ell) \to (\textit{coerce}\ B\ D\ \ell) \\
\textit{coerce}\ (\texttt{Code}\ T)\ (\texttt{Code}\ T)\ \ell &= \texttt{code-id}\ T \\
\textit{coerce}\ \texttt{Code}\star\ \texttt{Code}\star\ \ell &= \texttt{code-id}\star \\
\textit{coerce}\ \texttt{Code}\star\ (\texttt{Code}\ T)\ \ell &= \texttt{code?}^{\ell}\ T \\
\textit{coerce}\ (\texttt{Code}\ T)\ \texttt{Code}\star\ \ell &= \texttt{code!}\ T
\end{aligned}$$

**Figure 8.** The "coerce" function that generates coercions between types

We define a predicate $\mathbf{Empty^m}\ \Gamma$, which says that there is no metalanguage variable in the typing context $\Gamma$:

$$\frac{}{\mathbf{Empty^m}\ \emptyset} \qquad \frac{\mathbf{Empty^m}\ \Gamma}{\mathbf{Empty^m}\ (\Gamma, x{:}\texttt{otype}\ T)}$$

"Progress" says that a well-typed MetaCC term is either a value or a blame, or can take one step forward:

**Lemma 6.1** (Progress of MetaCC)**.** *If MetaCC term $M^c$ is well-typed: $\Gamma \vdash^c M^c : A$ and $\mathbf{Empty^m}\ \Gamma$, then*
*(1) $M^c$ is a value or*
*(2) $M^c$ is a blame: $M^c = \texttt{blame}\ \ell$ or*
*(3) $M^c$ can take a reduction step: $M^c \longrightarrow N^c$ for some $N^c$*

*Proof.* The proof is fully mechanized in `Progress.agda`. □

We then prove the small-step reduction of MetaCC preserves types:

**Lemma 6.2** (Preservation of MetaCC)**.** *If MetaCC term $M^c$ is well-typed: $\Gamma \vdash^c M^c : A$ and takes one reduction step to $N^c$: $M^c \longrightarrow N^c$, then $N^c$ is also well-typed: $\Gamma \vdash^c N^c : A$.*

*Proof.* The proof is fully mechanized in `Preservation.agda`. □

Multi-step reduction of MetaCC also preserves types:

**Lemma 6.3** (Multi-step reduction of MetaCC preserves types)**.** *If MetaCC term $M^c$ is well-typed: $\Gamma \vdash^c M^c : A$ and takes zero or more steps to $N^c$: $M^c \longrightarrow^* N^c$, then $N^c$ is also well-typed: $\Gamma \vdash^c N^c : A$.*

*Proof.* By induction on the multi-step reduction $M^c \longrightarrow^* N^c$. If it takes zero step, then $M^c$ is already well-typed. If it takes at least one step: $M^c \longrightarrow L^c$ and $L^c \longrightarrow^* N^c$ for some $L^c$, we apply Lemma 6.2 ("single step reduction preserves types") and then use the induction hypothesis. □

Our goal is to prove type safety for MetaGTLC. To connect MetaCC with MetaGTLC, we prove that the compilation from MetaGTLC to MetaCC preserves types:

**Lemma 6.4** (Compilation preserves types)**.** *If MetaGTLC term $M^m$ is well-typed:*

$$\Gamma \vdash^m M^m : A$$

*then the MetaCC term $M^c$ after compilation is also well-typed:*

$$\Gamma \vdash^m M^m : A \rightsquigarrow M^c \textit{ and } \Gamma \vdash^c M^c : A$$

*Proof.* The proof is fully mechanized in `CompilePres.agda`. □

We note that a value of type $(\texttt{Code}\ T)$ must be some quoted object language (STLC) code:

**Lemma 6.5** (Canonical form of quoted object code)**.** *If $\Gamma \vdash^c V : \texttt{Code}\ T$, then $V = \prec M^s \succ$ for some STLC term $M^s$.*

*Proof.* By inversion on the fact that $V$ is a value. Rule out other cases than the one for quote ($V = \prec M^s \succ$) by inversion on $\Gamma \vdash^c V : \texttt{Code}\ T$. □

We define well-typed metaevaluation results in Figure 10, which rule out `stuck`. Finally, we prove that metaevaluation always generates a well-typed result, which is a corollary of Lemma 6.3 and "compilation preserves types" (Lemma 6.4):

**Theorem 6.6** (Metaevaluation is type safe)**.** *If MetaGTLC term $M^m$ is well-typed:*

$$\emptyset \vdash^m M^m : \texttt{Code}\ T$$

*then the metaevaluation of $M^m$ generates a well-typed result:*

$$\vdash \textit{meta-eval}\ M^m : T$$

*Proof.* By Lemma 6.4, we have $\emptyset \vdash^m M^m : \texttt{Code}\ T \rightsquigarrow M^c$ and $\emptyset \vdash^c M^c : \texttt{Code}\ T$. By law of excluded middle, there are four cases: (1) $M^c$ reduces to a value, (2) $M^c$ reduces to a blame, (3) $M^c$ diverges, or (4) $M^c$ gets stuck.
(1) If $M^c$ reduces to some value $V$: $M^c \longrightarrow^* V$. Multi-step reduction preserves types (Lemma 6.3), so $\emptyset \vdash^c V : \texttt{Code}\ T$. By the canonical form of quoted object code (Lemma 6.5), $V = \prec M^s \succ$, so $\textit{meta-eval}\ M^m = M^s$ and $\vdash M^s : T$ (by rule *WT-STLC*).
(2) If $M^c$ reduces to a blame: $M^c \longrightarrow^* \texttt{blame}\ \ell$. The lemma is trivially true by rule *WT-blame*.
(3) If $M^c$ diverges: Trivially true by rule *WT-diverge*.
(4) Otherwise, $M^c$ gets stuck: $M^c \longrightarrow^* L^c$. $L^c$ is neither a value nor a blame, and there is no $N^c$ such that $L^c \longrightarrow N^c$, but that contradicts Lemma 6.1 ("progress"). □

$$\boxed{\Gamma \vdash^{o} M^{o} \Rightarrow T \rightsquigarrow M^{oc}}$$

$$\rightsquigarrow^{o}\textit{-const}\;\frac{k : \iota}{\Gamma \vdash^{o} k \Rightarrow \iota \rightsquigarrow k} \qquad \rightsquigarrow^{o}\textit{-var}\;\frac{\Gamma(x) = \mathtt{otype}\ T}{\Gamma \vdash^{o} x \Rightarrow T \rightsquigarrow x}$$

$$\boxed{\Gamma \vdash^{o} M^{o} \Leftarrow T \rightsquigarrow M^{oc}}$$

$$\rightsquigarrow^{o}\textit{-lam}\;\frac{(\Gamma, x{:}\mathtt{otype}\ S) \vdash^{o} M^{o} \Leftarrow T \rightsquigarrow M^{oc}}{\Gamma \vdash^{o} \lambda x.\, M^{o} \Leftarrow (S \to T) \rightsquigarrow \lambda x.\, M^{oc}}$$

$$\rightsquigarrow^{o}\textit{-app}\;\frac{\Gamma \vdash^{o} L^{o} \Rightarrow (S \to T) \rightsquigarrow L^{oc} \qquad \Gamma \vdash^{o} M^{o} \Leftarrow S \rightsquigarrow M^{oc}}{\Gamma \vdash^{o} L^{o}\ M^{o} \Rightarrow T \rightsquigarrow L^{oc}\ M^{oc}}$$

$$\rightsquigarrow^{o}\textit{-check-infer}\;\frac{\Gamma \vdash^{o} M^{o} \Rightarrow T \rightsquigarrow M^{oc}}{\Gamma \vdash^{o} M^{o} \Leftarrow T \rightsquigarrow M^{oc}}$$

$$\rightsquigarrow^{o}\textit{-ann}\;\frac{\Gamma \vdash^{o} M^{o} \Leftarrow T \rightsquigarrow M^{oc}}{\Gamma \vdash^{o} M^{o} : T \Rightarrow T \rightsquigarrow (M^{oc} : T)}$$

$$\rightsquigarrow^{o}\textit{-splice}\;\frac{\Gamma \vdash^{m} M^{m} : A \rightsquigarrow M^{c}}{\Gamma \vdash^{o} \sim^{\ell} M^{m} \Leftarrow T \rightsquigarrow \sim (M^{c}\langle \mathit{coerce}\ A\ (\mathsf{Code}\ T)\ \ell \rangle)}$$

$$\boxed{\Gamma \vdash^{m} M^{m} : A \rightsquigarrow M^{c}}$$

$$\rightsquigarrow^{m}\textit{-const}\;\frac{k : \iota}{\Gamma \vdash^{m} k : \iota \rightsquigarrow k} \qquad \rightsquigarrow^{m}\textit{-var}\;\frac{\Gamma(x) = \mathtt{mtype}\ A}{\Gamma \vdash^{m} x : A \rightsquigarrow x}$$

$$\rightsquigarrow^{m}\textit{-lam}\;\frac{(\Gamma, x{:}\mathtt{mtype}\ A) \vdash^{m} M^{m} : B \rightsquigarrow M^{c}}{\Gamma \vdash^{m} \lambda x{:}A.\, M^{m} : (A \to B) \rightsquigarrow \lambda x{:}A.\, M^{c}}$$

$$\rightsquigarrow^{m}\textit{-app}\;\frac{\begin{array}{c}\Gamma \vdash^{m} L^{m} : (A_1 \to A_2) \rightsquigarrow L^{c} \\ \Gamma \vdash^{m} M^{m} : B \rightsquigarrow M^{c} \qquad A_1 \sim B\end{array}}{\Gamma \vdash^{m} (L^{m}\ M^{m})^{\ell} : A_2 \rightsquigarrow L^{c}\ (M^{c}\langle \mathit{coerce}\ B\ A_1\ \ell \rangle)}$$

$$\rightsquigarrow^{m}\textit{-app}\star\;\frac{\Gamma \vdash^{m} L^{m} : \star \rightsquigarrow L^{c} \qquad \Gamma \vdash^{m} M^{m} : A \rightsquigarrow M^{c}}{\Gamma \vdash^{m} (L^{m}\ M^{m})^{\ell} : \star \rightsquigarrow (L^{c}\langle \mathit{coerce}\ \star\ (\star \to \star)\ \ell \rangle)\ (M^{c}\langle \mathit{coerce}\ A\ \star\ \ell \rangle)}$$

$$\rightsquigarrow^{m}\textit{-quote}\;\frac{\Gamma \vdash^{o} M^{o} \Rightarrow T \rightsquigarrow M^{oc}}{\Gamma \vdash^{m} \prec M^{o} \succ : \mathsf{Code}\ T \rightsquigarrow \prec M^{oc} \succ}$$

**Figure 9.** Compilation rules from MetaGTLC to MetaCC

$$\boxed{\vdash r : T}$$

$$\textit{WT-STLC}\;\frac{\emptyset \vdash^{oc} M^{s} : T}{\vdash M^{s} : T} \qquad \textit{WT-blame}\;\frac{}{\vdash \mathtt{blame}\ \ell : T}$$

$$\textit{WT-diverge}\;\frac{}{\vdash \mathtt{diverge} : T}$$

**Figure 10.** Well-typed metaevaluation results

## 7 Conclusion

In this paper, we proposed an improvement to the error detection of dynamically-typed DSLs that generate statically-typed data pipeline descriptions. We introduced gradual metaprogramming, which incrementally type checks code fragments as they are spliced together and reports the source location of the problem when type checking fails. Gradual metaprogramming provides three main benefits: first, it provides a migration path toward statically-typed DSLs. Second, it enables earlier error detection. Last but not the least, it pinpoints the source code location responsible for the type error in the metaprogram. We defined MetaGTLC, a metaprogramming calculus in which a gradually-typed metalanguage manipulates a statically-typed object language. We proved type safety for MetaGTLC, which says that successful metaevaluation always generates well-typed object code. We mechanized MetaGTLC and the type safety proof in Agda.